\documentclass[aps,prd,twocolumn,showpacs,article,floatfix]{revtex4-1}
\pdfoutput=1 

\usepackage{fullpage}
\usepackage{amsmath}
\usepackage{amssymb}
\usepackage{graphicx}

\begin{document}

\title{An Exact Model of the Power/Efficiency Trade-Off\\ While Approaching  the Carnot Limit}
\author{Clifford V. Johnson}
\email{johnson1@usc.edu}
\affiliation{Department of Physics and Astronomy\\ University of
Southern California \\
 Los Angeles, CA 90089-0484, U.S.A.}

\pacs{05.70.Ce,05.70.Fh,04.70.Dy}

\begin{abstract}
The Carnot heat engine sets an upper bound on the efficiency of a heat engine. As  an ideal, reversible  engine, a  single  cycle must be performed in infinite time,  and so the Carnot engine has zero power. However, there is nothing in principle forbidding the existence of a heat engine whose efficiency approaches that of Carnot while maintaining finite power. Such an engine must have very special properties, some of which have been  discussed in the literature, in various limits. While recent theorems rule out a large class of engines from  maintaining finite power at exactly the Carnot efficiency, the approach to the limit still merits close study. Presented here is an exactly solvable model of such an approach that may serve as a laboratory for  exploration of the underlying mechanisms. The equations of state have their origins in the extended thermodynamics of electrically charged black holes.
\end{abstract}

\keywords{wcwececwc ; wecwcecwc}

\maketitle

It is well known that a heat engine, regardless of working substance and the details of the thermodynamic cycle, has a fundamental limit on its efficiency given by the Carnot efficiency $\eta_{\rm C}^{\phantom{C}}$. If $T_H$  is the highest operating temperature in the engine and $T_C$ the lowest,  (the temperatures at which the input heat  $Q_H$ and exhaust heat $Q_C$ are exchanged with the hot and cold reservoirs, respectively), the efficiency~$\eta$ is bounded as follows: 
\begin{equation}
\eta=1-\frac{Q_C}{Q_H} \leq \eta_{\rm C}^{\phantom{C}}=1-\frac{T_C}{T_H}\ .
\end{equation}
It is also familiar that the Carnot engine itself is an idealized reversible engine, with a cycle that is composed of two isotherms and two adiabats, with the expansions and compressions performed quasi--statically, in order to maintain reversibility. In other words, it takes an infinite amount of time to perform one cycle of the Carnot engine: It has zero power. 

While most typical heat engines, working at finite power, operate well below the Carnot efficiency, there is no issue of principle that prevents their efficiency from approaching that of Carnot, but it becomes increasingly difficult for typical working substances and choices of operating cycle.  The question naturally arises as to what kind of engine is needed to approach the Carnot efficiency while maintaining finite power. (This is a separate issue from the Curzon--Ahlborn bound on $\eta$ when working at maximum power  \cite{1975AmJPh..43...22C}.) There have been  recent discussions of this in the  thermodynamics and statistical mechanics literature  \cite{2011PhRvL.106w0602B,2013PhRvL.111e0601A,2015PhRvL.115i0601P,PhysRevX.5.031019,2015PhRvL.114e0601P,2016NatCo...711895C,2016EPJB...89..248K,2016PhRvL.117s0601S,2017arXiv170101914S}, and two papers in particular  \cite{2015PhRvL.114e0601P,2016NatCo...711895C} consider having the working substance near criticality as an  approach to the problem,  exploiting either fluctuations, or a diverging heat capacity to argue for  the maintenance of finite power as  $\eta$ grows closer to $\eta_{\rm C}^{\phantom{C}}$. It has been argued in refs. \cite{2016PhRvL.117s0601S,2017arXiv170101914S}   that it is forbidden (for widely applicable assumptions) to be exactly at the Carnot efficiency while at finite power, but the issue of the approach to the limit is still of considerable interest, for both practical and theoretical reasons. This paper presents an exactly solvable model of such an approach that may be of use in gaining better understanding of how various models (perhaps less computationally accessible) work. A critical system will also feature in the present work, but its role appears to be of quite a different character from  what was argued for in the  systems of refs. \cite{2015PhRvL.114e0601P,2016NatCo...711895C}. Fluctuations and diverging specific heat do not explicitly play an essential role in the core construction. This can be determined because the system employed is can be readily queried with a computation: All the needed properties of the working substance are available {\it via} a full set of exact defining equations.

The system to be used here has its origins in  {\it extended}  gravitational thermodynamics: The traditional  treatment % \cite{Bekenstein:1973ur,Bekenstein:1974ax,Hawking:1974sw,Hawking:1976de} 
 \cite{Gibbons:1976ue} of black holes  in semi--classical quantum gravity supplies them with a temperature~$T$ and an entropy $S$, which depend upon the mass~$M$ and the horizon radius $r_+$. This treatment is extended     \cite{Kastor:2009wy} by  making dynamical the cosmological constant ($\Lambda$)  of the gravity theory \footnote{A dynamical $\Lambda$ can be naturally implemented if the gravity theory is embedded within a theory with other dynamical sectors. An example is when there are dynamical scalars $\varphi_i$ with a potential $V(\varphi_i)$. Moving between fixed points of the potential, where the scalars take on fixed values $\varphi^{\rm c}_i$, that set a non--zero constant $V(\varphi^{\rm c}_i)$,  setting the value of $\Lambda$. See {\it e.g.} the review in ref. \cite{Aharony:1999ti}.}, which yields a pressure variable  $p=-\Lambda/8\pi$ \footnote{Here we are using geometrical units where $G,c,\hbar,k_{\rm B}$ have been set to unity.} and its conjugate volume~$V\equiv (\partial H/\partial p)_S$. The enthalpy $H$ is the black hole's mass,
and the First Law in terms of all these quantities is $dH=TdS+Vdp$. Studies of the extended thermodynamics of gravitational systems have uncovered many phenomena that are familiar from statistical physics and thermodynamics. (For a recent review see ref. \cite{Kubiznak:2016qmn}.)

Notice that for negative $\Lambda$ the pressure is positive. Ref. \cite{Johnson:2014yja} presented the idea of defining heat engines that do traditional mechanical work $W=\int pdV$ in this context \footnote{Work can be interpreted here results in a change of the  overall energy of a spacetime since the volume removes removes portions of it from the standard energy integral. Recall that $p$ sets an energy density {\it via} $\Lambda$'s equation of state $\rho=-p$ and so a positive change $dV$ results in an energy gain $|\rho| dV$. See ref. \cite{Kastor:2009wy}.}. The heat flows $Q_H$ and $Q_C$ into and out of the engine can be considered as from and to non--backreacting heat baths of radiation  filling the spacetime, as is traditional in black hole thermodynamics. (See {\it e.g.} ref. \cite{Hawking:1982dh}.)  Such engines were called holographic heat engines since gravitational physics in spacetimes with negative $\Lambda$ is known to have a dual description in terms of strongly coupled non--gravitational physics (in one dimension fewer). % \cite{Maldacena:1997re,Witten:1998qj,Gubser:1998bc,Witten:1998zw,}. 
These are examples of a broader phenomenon in quantum gravity known as holography. (For a review, see ref. \cite{Aharony:1999ti}). Such dualities are not needed here, but it is worth noting that if the gravitational language is not to a reader's taste, this could all in principle be translated to the language of a class of strongly coupled  gauge theories. In other words, the gravitational aspect of this example is not essential, but it is more economical to use that simpler language.

The context will be a $(3{+}1)$--dimensional  Einstein--Maxwell system  with action:
\begin{equation}
I=-\frac{1}{16\pi }\int \! d^4x \sqrt{-g} \left(R-2\Lambda -F^2\right)\ ,
\label{eq:action}
\end{equation}
 where $\Lambda{=}-3/l^2$  sets a length scale $l$. The black hole  spacetimes of interest here are Reissner--Nordstrom--like solutions of charge~$q$.  The metric and gauge potential are:
\begin{eqnarray}
ds^2 &=& -Y( r)dt^2
+ {dr^2\over Y(r)} + r^2 (d\theta^2+\sin^2\theta d\varphi^2)\ ,\nonumber \\
Y( r) &\equiv& 1-\frac{2M}{r}+\frac{q^2}{r^2}+\frac{r^2}{l^2}\ , A_t = q\left(\frac{1}{r_+}-\frac{1}{r}\right)\! .
\end{eqnarray}
%and  $A_t = q(r-r_+)/{rr_+}$. 
The potential  is chosen to vanish on the horizon  at $r=r_+$, the largest positive real root of $Y(r)$.

 Several aspects of the thermodynamics of these solutions were studied in refs. \cite{Chamblin:1999tk,Chamblin:1999hg}. There, a rich phase structure was uncovered, a  van der Waals--like nature was elucidated, including a second order critical point. By including   variable $\Lambda$ (and hence  a pressure $p$), ref. \cite{Kubiznak:2012wp} clarified the resemblance to van der Waals and showed that the system has the same universal behaviour near the critical point as the van der Waals gas.  

\begin{figure}[h]
 %\begin{wrapfigure}{l}{0.3\textwidth}
%\begin{center}
%{\centering
\includegraphics[width=3.0in]{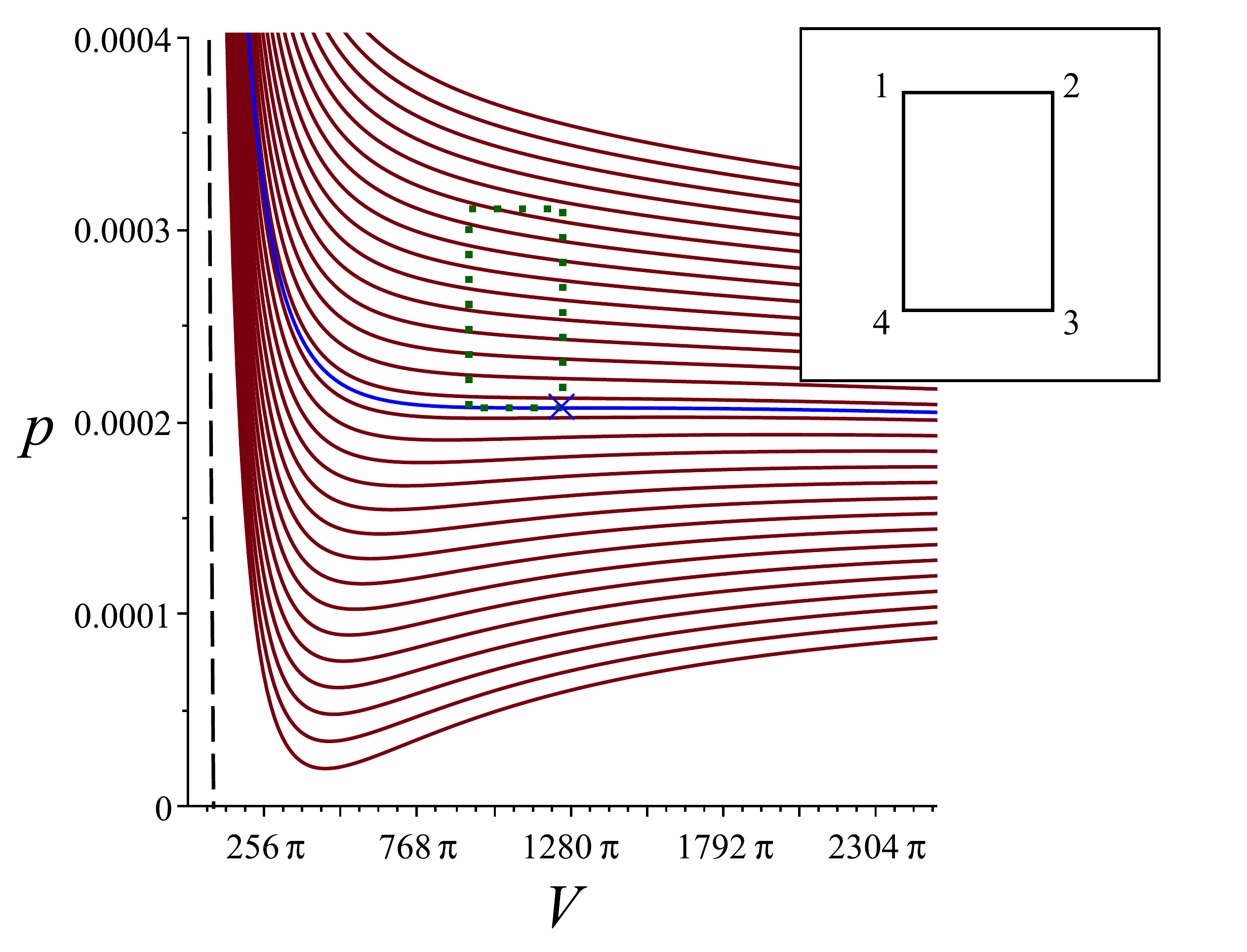} 
   \caption{\footnotesize  Main: Sample isotherms for $q{=}4$. The temperature is higher for the curves further away from the origin. The central (blue) isotherm is at the critical temperature, and the (blue) cross marks the critical point. The isotherms at lower temperatures get modified, as discussed in the main text, but this is not shown here. The dotted green rectangle is an example of the   special  engine cycle discussed  in the text (with $L{=}1$). The dashed curve is the $T{=}0$ isotherm. Inset: The labelling of the engine cycle.}   \label{fig:isotherms}
%}
%   \end{center}
%\end{wrapfigure}
\end{figure}

%Key to all this is that $\it via$ the standard \cite{Bekenstein:1973ur,Bekenstein:1974ax,Hawking:1974sw,Hawking:1976de} semi--classical quantum gravity procedures ({\it e.g.,} requiring smoothness at the horizon in the Euclidean section) 
The standard semi--classical quantum gravity procedures  \cite{Gibbons:1976ue}  yield  a temperature for each black hole solution, which depends on $r_+$, $q$, and $\Lambda$.  The entropy is given by a quarter of the area of the horizon: $S=\pi r_+^2$. The extended thermodynamics \cite{Kastor:2009wy} allows for all appearances of the length scale~$l$ to be replaced by the pressure $p$ using the relation $p=3/(8\pi l^2)$, and the thermodynamic volume turns out to be $V=4\pi r_+^3/3$.  So all occurrences of $r_+$ can be traded in for either  an $S$ or a~$V$, as they are not independent.  All of this results in an equation of state $p(V,T)$:
\begin{equation}
%p=\frac{T}{v}-\frac{1}{2\pi v^2}+\frac{2q^2}{\pi v^4}
p=\frac{1}{8\pi}\left(\frac{4\pi}{3}\right)^\frac43\left(\frac{3T}{V^\frac13} - \left(\frac{3}{4\pi}\right)^{\frac23}\frac{1}{V^\frac23}+\frac{q^2}{V^\frac43}\right)\ .
\label{eq:equationofstate}
\end{equation}  
Some sample isotherms are plotted in figure~\ref{fig:isotherms}. Note that there is a wedge--shaped exclusion region extending from the $V=0$ axis, bounded on the right by the $T=0$ curve (the dashed line) and on the bottom by the $p=0$ line. Points inside that region are unphysical, having $T<0$. (The black hole at $(T=0, p=0)$ is the extremal Reissner--Nordstrom black hole of volume $V_0=4\pi q^3/3$.) Below a critical isotherm  the isotherms yield multiple values for $p$, and (in full parallel with the classic van der Waals system) are ``repaired" by an isobar (not shown in figure~\ref{fig:isotherms}) at a value of the pressure determined by a study of the free energy. This results in a family  of first order phase transitions between large and small black holes terminating in a second order critical point at the critical isotherm \cite{Chamblin:1999tk,Chamblin:1999hg}. The details of the first order transitions  will not affect the main issue being addressed here, and so they won't be explored further.

An equivalent expression to eq.~(\ref{eq:equationofstate}) is:
\begin{equation}
T=\frac{1}{4\sqrt{\pi}}S^{-\frac32}\left(8pS^2+S-{\pi q^2}\right)\ .
\label{eq:temperature}
\end{equation}
Meanwhile the mass (enthalpy) $H(S,p)$ is  \cite{Dolan:2011xt}:
\begin{equation}
M\equiv H=\frac{1}{6\sqrt{\pi}} S^{-\frac12}\left(8pS^2+3S+3\pi q^2\right) \ ,
\label{eq:enthalpy}
\end{equation}
and  the constant $V$ and $p$ specific heats are  \cite{Kubiznak:2012wp}:
\begin{equation}
C_V=0 \ ;  C_p=2S\left(\frac{8pS^2+S-\pi q^2}{8pS^2-S+3\pi q^2}\right)\ .
 \end{equation}

 It is these black holes that were the working substance in the prototype holographic heat engine of ref.~\cite{Johnson:2014yja}, using a rectangular cycle in the $(p,V)$--plane made of isobars and adiabats (which are equivalent to isochors for static black holes since $S$ and $V$ both depend only on $r_+$).  The inset of figure~\ref{fig:isotherms} shows the labelling of the cycle  to be used below.  
%\begin{figure}[h]
% %\begin{wrapfigure}{l}{0.3\textwidth}
%%\begin{center}
%%{\centering
%\includegraphics[width=1.5in]{special_{\rm cr}ycle} 
%   \caption{\footnotesize  The prototype engine cycle.}   \label{fig:cyclesb}
%%}
%%   \end{center}
%%\end{wrapfigure}
%\end{figure} 
Later, in ref.~\cite{Johnson:2016pfa}, an exact  equation for the efficiency was derived for the cycle, and it will be extremely  useful here. Key is that the heat flows can be written as mass/enthalpy differences, giving:
\begin{equation}
\eta=1-\frac{Q_C}{Q_H}=1-\frac{M_3-M_4}{M_2-M_1}\ ,
\label{eq:efficiency-prototype}
\end{equation}
where $M_i$ is the black hole mass evaluated at the $i$th corner. Its simplicity means that there is no need to make the kinds of approximations ({\it e.g.} high temperature or pressure) usually needed to write explicit efficiency formulae for some particular choice of location of this cycle in the $(p,V)$ plane.  

The next step is to decide where to place the cycle in the $(p,V)$ plane. In  previous work in this area, $q$ has been treated essentially as a label for a family of solutions, and was conveniently set to a positive non--zero value and forgotten about, since the key features don't depend upon its actual value. This will not be the case here. Consider making $q$ large, for reasons that will become clear shortly. For a given choice of the position of the cycle (choosing a range for $p$ and~$V$ (or $S$)), a sensible engine can be defined, but for large enough $q$ eventually the system will become unphysical:  $T$ (see  eq.~(\ref{eq:temperature})) on some parts of the cycle starts becoming negative because the exclusion region moves to the right with increasing $q$. This can be avoided by  seeking choices for the $p,V$ (or $S$) coordinates of the cycle  variables that scale with $q$ in such a way as to stay physical. Eq.~(\ref{eq:temperature})  shows that the scaling is $S\sim q^2$, $p\sim q^{-2}$, and hence $T\sim q^{-1}$. There is a very distinguished point exhibiting  this exact scaling behaviour. It is the second order critical point, defined by the $p(V,T)$ curve with a point of inflection: $\partial p/\partial V=\partial^2 p /\partial V^2=0$:
\begin{equation}
\label{eq:critical}
p_{\rm cr}=\frac{1}{96\pi q^2}\ , \quad S_{\rm cr}=6\pi q^2\ , \quad T_{\rm cr}= \frac{1}{3\sqrt{6}\pi q}\ ,
\end{equation}
with $V_{\rm cr}= 8\sqrt{6}\pi q^3.$ (See fig.~\ref{fig:isotherms} for the case of $q=4$.)

% \cite{,Caldarelli:1999xj,Wang:2006eb,Sekiwa:2006qj,LarranagaRubio:2007ut,Dolan:2010ha,Cvetic:2010jb,Dolan:2011jm,Dolan:2011xt,Henneaux:1984ji,Teitelboim:1985dp,Henneaux:1989zc}). 

So if  the engine cycle is chosen to be in the neighbourhood of this point, and of a size that does not extend into the exclusion region, it will be physical. There are several ways of making such a choice, and one family will be chosen here for illustration.  Place  the critical point at corner 3: $p_3=p_4=p_{\rm cr}$, and choose the upper isobar \footnote{The  choice of placing the critical point on the top isobar can also be made. The cycle  dips into the repaired region  where there are first order transitions. However, it is easier to determine the temperatures on the isobars away from that region, and so for the sake of simplicity, the bottom isobar was chosen.}  to be at some multiple of $p_{\rm cr}$: $p_1=p_2=3p_{\rm cr}/2$. (See the dotted rectangle in fig.~\ref{fig:isotherms}.) In preparation for large $q$ it is to be noted that since  $p\sim q^{-2}$, the cycle is in danger of shrinking to zero area, giving vanishing work and hence  vanishing $\eta$. But if  $V_2-V_1=V_3-V_4$ are chosen to scale as $q^2$, then the work will be finite at any $q$. So $V_2=V_3=V_{\rm cr}$ while $V_1=V_4= V_{\rm cr}-V_{\rm cr}L/q$, where $L$ is a constant. This gives $W=p_{\rm cr}V_{\rm cr}L/2q=L/4\sqrt{6}$.  So as $q$ is increased, the whole cycle shrinks vertically, but grows horizontally in such  a way as to keep the work done finite. It is now a matter of studying the $q$ dependence of the input heat $Q_H$.  It is simply the mass (enthalpy) difference $M_2-M_1$, with $p=3p_{\rm cr}/2$, $S_2=S_{\rm cr}$ and $S_1=S_{\rm cr}(1-L/q)^\frac23$ placed into eq.~(\ref{eq:enthalpy}).
%\begin{equation}
%Q_H= \frac{2\sqrt{6}}{3}q - \frac{1}{6\sqrt{\pi}} S_1^{-\frac12}\left(8p_{\rm cr}S_1^2+3S_1+3\pi q^2\right)\ ,
%\end{equation}
%with . 
Interestingly, the large $q$ expansion of  $Q_H$ decreases to a limiting value:
\begin{equation}
Q_H= \frac{19\sqrt {6}}{72}L+\frac{\sqrt {6}}{27}{\frac {{L}^{2}}{q}}+{\frac {4\,{L}^{3}
\sqrt {6}}{243\,{q}^{2}}}+O \left( {q}^{-3} \right)\ , %+{\frac {25\,{L}^{4}\sqrt {6}}{2916\,{q}^{3}}
%}+O \left( {q}^{-4} \right) 
\end{equation}
and hence the efficiency $\eta=W/Q_H$ is, at large $q$:
\begin{equation}
\eta=\frac{3}{19}-\frac{8}{361}{\frac {L}{q}}-{\frac {416\,{L}^{2}}{61731\,{q}^{2}}}-{\frac {3286{
L}^{3}}{1172889\,{q}^{3}}}+O \left( {q}^{-4} \right) \ .
\end{equation}
The next thing to do is compute the Carnot efficiency for the engine. Directly inserting the chosen values  for ($S_2,p_2$) and ($S_4,p_4$) into eq. (\ref{eq:temperature}) gives $T_H{=}(19\sqrt{6}/288)(\pi q)^{-1}$ exactly, while  the large $q$ expansion for $T_C$ begins:
\begin{equation}
T_C= \frac{1}{18}\,{\frac {\sqrt {6}}{\pi\,q}}-{\frac {{L}^{3}\sqrt {6}}{972\,\pi\,
{q}^{4}}}-{\frac {7\,{L}^{4}\sqrt {6}}{3888\,\pi\,{q}^{5}}}+O \left( {
q}^{-6} \right) \ ,
\end{equation}
 and so:
\begin{equation}
\label{eq:carnot_critical}
\eta_{\rm C}^{\phantom{C}}=1-\frac{T_C}{T_H} = {\frac{3}{19}}+{\frac {8\,{L}^{3}}{513\,{q}^{3}}}+{\frac {14\,{L}^{4}
}{513\,{q}^{4}}}+O \left( {q}^{-
5} \right) 
\ .
\end{equation}

\begin{figure}[h]
 %\begin{wrapfigure}{l}{0.3\textwidth}
%\begin{center}
%{\centering
\includegraphics[width=2.0in]{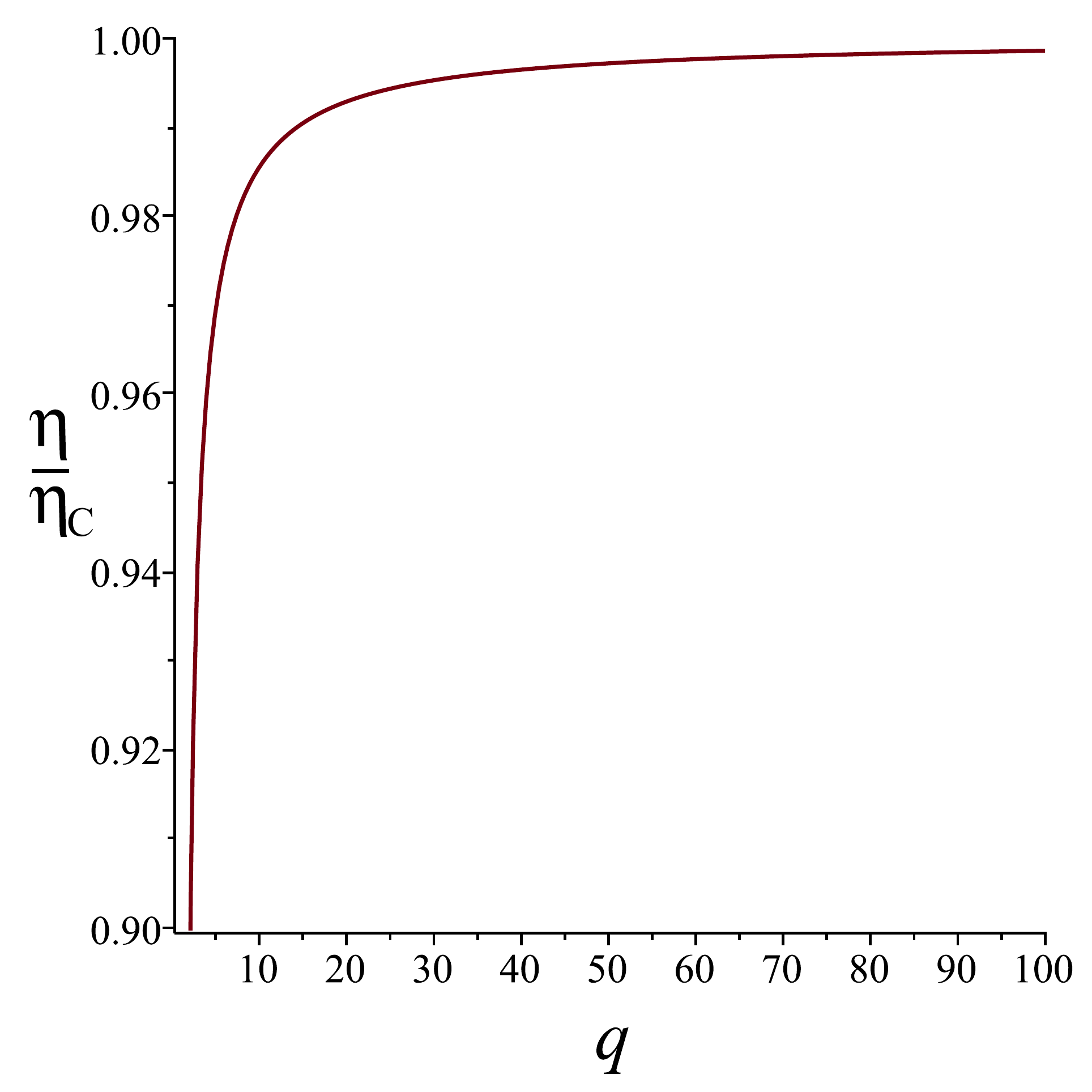} 
   \caption{\footnotesize  The behaviour of the ratio $\eta/\eta_{\rm C}$ as a function of $q$ for the prototype scheme. Variations of the scheme discussed in the text behave qualitatively similarly.}   \label{fig:ratio}
%}
%   \end{center}
%\end{wrapfigure}
\end{figure}

These  simple but striking results constitute the main demonstration promised for this paper. (Fig.~\ref{fig:ratio} is a plot of the ratio $\eta/\eta_{\rm C}^{\phantom{C}}$ {\it vs.} $q$, showing the rise to unity at large $q$.) This is a heat engine that does finite work at any $q$, and $\eta\to\eta_{\rm C}^{\phantom{C}}$ as $q\to\infty$. This is atypical, since usually going to the  Carnot limit for one of the classic heat engine cycles (or variants thereof)  translates into vanishing or infinite work.  As an  example, the Otto cycle has $\eta=1-r^{1-\gamma}$, where $\gamma=C_p/C_V>1$ and $r$ is the ratio of largest to smallest volumes, and so $\eta$ is maximized for $r\to\infty$. As another, the (Brayton--like)  rectangular cycle defined for black holes at  high pressures and temperatures well away from the critical point has \cite{Johnson:2014yja} $\eta=(1-p_4/p_1)(1+O(1/p_1)\cdots)$ (using the labelling in figure~\ref{fig:isotherms}), which may be written as:
\begin{equation}
\eta=1-\frac{T_C}{T_H}\left(\frac{V_2}{V_4}\right)^{1/3}+\cdots \ ,
\end{equation}
This is an analogue of an ideal gas limit \cite{Johnson:2015ekr}, and Carnot efficiency is approached if $V_2\to V_4$, resulting in no work. In the case under consideration however, the region of interest is far from an ideal gas regime and in fact as $q$ grows $T$ decreases. This is appealing since most of the current discussions in the literature about physical realizations of finite power efficient engines are about low temperature experiments. Finite work here as $\eta\to\eta_{\rm C}^{\phantom{C}}$ is a useful feature to have under control on the way to studying finite power.  

The time taken to do a cycle, $\tau$, is all that needs to be examined next. On the face of it, that seems to be finite at any finite $q$ (but see further discussion  below) and so this is indeed a model of an {\it approach} to Carnot while maintaining finite power in the following precise sense: An efficiency as close to the Carnot efficiency as desired can be achieved by choosing large enough~$q$, and picking the cycle according to the prescription above. Precisely at the limit $\eta=\eta_{\rm C}^{\phantom{C}}$, however, while the work is finite, the power has vanished, since the pressures in the isobars are proportional to $p_{\rm cr}\sim q^{-2}$, which vanishes in the limit, meaning that the time it takes to perform the isobaric expansions and compressions diverges. So the inequality of {\it e.g.} refs.\cite{2016PhRvL.117s0601S,2017arXiv170101914S} showing the unattainability of finite power exactly at Carnot efficiency is easily satisfied. It is:
\begin{equation}
\frac{W}{\tau}\leq{\bar\Theta} \frac{\eta(\eta_{\rm C}^{\phantom{C}}-\eta)}{T_C}\ ,
\end{equation}
where ${\bar\Theta}$ is a model dependent constant capturing the characteristics of the engine. For the current model, the right hand side (divided by ${\bar\Theta}$) has a large $q$ expansion that begins as: $%\eta(\eta_{\rm C}-\eta)/T_C= 
{ {72\pi\sqrt {6}}L/{6859}}+{ {224\pi\sqrt {6}}L^2/{(130321
q)}}%+{\frac {19358\,\pi\,\sqrt {6}}{2476099\,{q}^{2}}}
+O \left( {q}^{-2}\right).$ Meanwhile, the work $W{=}L/4\sqrt{6}$, but an optimistic estimate of the~$q$ dependence of $\tau$ based on the behaviour of the pressures (discussed above) is $\tau\sim q^2$.

The above cycle  is just one example of the kind of scheme that will work.  Variations  were studied, and some are worth reporting the results for. One  is to  put the critical point at a different point on the lower isobar. This results in qualitatively similar results at large $q$. The difference is that both  $T_H$ and $T_C$ have higher order corrections to their leading $1/q$ form at large $q$. Again, $\eta$ and $\eta_{\rm C}^{\phantom{C}}$ converge at large~$q$ to $3/19$. It should also be noted that the approach $\eta\to\eta_{\rm C}$ at large $q$ is achieved even if the critical point is not anywhere on the cycle itself. It suffices to be near enough to the critical region, in the manner outlined.  %More details will appear in a longer publication  \cite{metoappear}.

A concern that might be raised is whether the presence of  a critical point somewhere on the cycle might invalidate the claim to be able to achieve finite power (at large but finite $q$) due to possible critical slowing of the system. As mentioned above, similar results were  achieved by avoiding the critical point, only having the cycle be near it,  with $\eta$  approaching $\eta_{\rm C}^{\phantom{C}}$ at large $q$ as before. So  the critical point's presence directly on the cycle is not crucial. There might have been an expectation that certain aspects of  the physics near the critical point itself are responsible for the (finite power) approach to the Carnot efficiency at large $q$. For example, in the discussion in ref. \cite{2016NatCo...711895C} it is argued that the divergence of the specific heat produces an enhancement in the power's scaling with  the effective system size, $N$, through an enhancement in the heat flows. (There are $N$ coupled quantum Otto engines constituting the system.) Here,~$q$ acts as a system size parameter analogous to that paper's $N$. Indeed, near criticality  $C_p$'s peak (inverse) width and height are 
%\cite{metoappear} 
enhanced with increasing  $q$, as happens in ref. \cite{2016NatCo...711895C}. But the explicit expressions for $Q_H$ and $Q_C{=}W{-}Q_H$ show that they actually decrease with increasing~$q$, even with the critical point on the isobar. Also, the fact that the same qualitative behaviour happens away from the critical point suggests that in this model the peak in $C_p$ plays no crucial role in driving the efficiency toward Carnot. On the other hand, since the construction presented here removes the $q$ dependence (analogously, $N$ dependence) from the work and places it all into the heat flows, it is difficult to compare the approaches further.

Nevertheless, the critical point itself {\it is} important in the whole scheme, since as shown, its neigbourhood (which depends on $q$, see eq.~(\ref{eq:critical})) is key in determining the coordinates of the cycle needed to approach the Carnot efficiency as $q$ increases. A qualitative reason why this all works so well is as follows: The neighbourhood of the critical point on the critical isotherm, being a region containing a point of inflection, is locally quite horizontal. Other isotherms in the region will inherit some of this behaviour, and this is even more true at higher $q$. Close to horizontal means that they do not deviate too far from the isobar shape of the 1--2 and 3--4 parts of the cycle. As discussed earlier, the vertical 2--3 and 4--1 isochoric parts are also adiabats (because of the properties of static black holes). So a prescription for picking a cycle that stays in the neighbourhood of the critical point therefore ensures that the cycle itself becomes an increasingly better approximation to a Carnot cycle (two isotherms and two adiabats) as $q$ grows. The behaviour of the pressures resulting from this is such that they will vanish in the limit and result in diverging $\tau$ as expected for Carnot.

The underlying system controlling the physics at large $q$ is worth further investigation: It has low pressure and temperature, and high volume and entropy. In the gravitational  model it originates as a special family of large charge black holes, but there might be analogues of such equations of state in other, non--gravitational, systems. They would be interesting to identify.

\medskip

 \begin{acknowledgments}
CVJ  thanks the  US Department of Energy for support under grant DE-FG03-84ER-40168,  the Simons Foundation for  a Simons Fellowship (2017), and Amelia for her support and patience.    
\end{acknowledgments}

\bibliographystyle{apsrev4-1}
\bibliography{johnson_carnot}

\end{document}